\documentclass[11pt,citesort]{article}
\usepackage{amsmath}
\usepackage{amsfonts}
\usepackage{amssymb}
\usepackage{graphicx}
\usepackage{subfigure}

\usepackage{color}

\newcommand\be{\begin{equation}}
\newcommand\ee{\end{equation}}
\newcommand\bea{\begin{eqnarray}}
\newcommand\eea{\end{eqnarray}}
\newcommand{\bdm}{\begin{displaymath}}
\newcommand{\edm}{\end{displaymath}}
\newcommand\nn{ \nonumber\\}

\newcommand{\footnoteremember}[2]{  \footnote{#2}\newcounter{#1} \setcounter{#1}{\value{footnote}}} 
\newcommand{\footnoterecall}[1]{  \footnotemark[\value{#1}]}

\title{BCFW Deformation and Regge Limit}

\author{ Chih-Hao Fu,\footnoteremember{NCTU}{Department of Electrophysics, National Chiao-Tung University and Physics Division, National Center for Theoretical Sciences, Hsinchu, Taiwan, R.O.C.}\\
Jen-Chi Lee,\footnoterecall{NCTU}\\
Chung-I Tan,\footnoteremember{Brown}{Physics Department, Brown University,
Providence, RI 02912, USA}\\
and\\
Yi Yang,\footnoterecall{NCTU}
}

\begin{document}
 
\maketitle

\begin{abstract}
BCFW deformation has served as an extremely useful tool in providing a recursive approach in studying color-ordered gauge amplitudes.  This procedure  has also been generalized to the study of graviton scattering.  An important ingredient of this approach is the ability to identify amplitudes satisfying convergent  dispersion relation when the BCFW parameter, $z$, is treated as a complex variable.  In a {\it modified} BCFW treatment, we show in what sense the BCFW deformation in the large-z limit can be understood as the Regge limit. We also discuss how the issue of convergent dispersion integral for amplitudes involving external spins relates to the study of super-convergence relation which served as the precursor to the s-t duality relation for flat space string amplitudes. 
 \end{abstract}

\newpage
\setcounter{tocdepth}{3}

\newpage
\section{Introduction}
\label{sec:intro}

Rapid progress has been made over the past decade in understanding the structure of gauge-amplitudes through the use of spinor calculus. Increasingly in use is the exploration of analyticity and unitarity constraints~\cite{ArkaniHamed:2008gz}. This is particularly useful in the large-N limit where amplitudes can be expressed as a sum over color-orderings, and, for each color-ordered amplitude, it admits a topological expansion, with the leading order being planar.  An important development  has been the use of BCFW deformation \cite{Britto:2004ap,Britto:2005fq,ArkaniHamed:2008yf}. For each colored ordered amplitude, $A(p_1, p_2, \cdots, p_n)$,  consider shifting  a  pair of momenta $(p_i,p_j)$ where  $p_i \rightarrow  \hat p_i(z)=p_i+z q$ and $ p_j \rightarrow  \hat p_j(z) =p_j-zq\; , $
with $q^2=q\cdot p_i=q\cdot p_j=0$. By treating  the shifted amplitude $A_n(z) \equiv A(\cdots \hat p_i \cdots \hat p_j \cdots)$  as an analytic function of $z$, a robust recursive approach for  calculating multi-gluon amplitudes has been developed through the use of dispersion relations.\footnote{See ref. \cite{Feng:2011np} for a recent review and references therein.}  More generally, a key requirement for a dispersive approach  is  the control on the large-$z$ behavior of the shifted amplitude, $A_n(z)$.     It has also been emphasized~\cite{ArkaniHamed:2008gz} that the existence of  ``unsubtracted dispersion relation" (UDR)  holds only for theories involving spins, e.g., it holds for gauge theories and gravity, but fails for scalar theories, e.g., 
$\lambda \phi^4$.
In cases where the large-$z$ asymptotic fails to converge, the corresponding boundary contribution can
be separately calculated and subtracted from the
dispersion relation~\cite{Feng:2010ku,Feng:2009ei}. Alternatively, it can be translated into zeros,
from which a modified relation can be derived~\cite{Benincasa:2011kn,Benincasa:2011pg}. 
Recent discussions on large-$z$ behavior  can
be found in~\cite{Boels:2011tp,Boels:2011mn,Britto:2012qi,Zhang:2013cha,Chen:2012vz,Boels:2012sy,Du:2011se,Du:2011fc}.

Kinematically, the large-$z$ limit of the BCFW deformation  can be thought of  as a Regge limit~\footnote{A pedagogical review of Regge theory is provided in Appendix A. For a historical backdrop leading to the development of the Regge theory, see \cite{Chew,Gribov,DTFT}.}. For example,  consider a 4-point amplitude, $A(s,t)$,  expressed in terms of Mandelstam invariants.  Under BCFW deformation for the adjacent pair $(p_1,p_2)$, we have 
\be
s\rightarrow s(z)\equiv 2\hat p_2(z)\cdot p_3 = s(0) + b \; z
\ee
while $t=2 \hat p_1\cdot \hat p_2= 2  p_1\cdot p_2$ remains unshifted. Therefore, the limit of  large $z$  is formally the same  as the Regge limit  of large $|s|$, with $t$ fixed, 
$ \lim_{s(z)\rightarrow \infty}  A(s(z), t) $.  In this paper, we focus on  developing this connection.   If the equivalence between the BCFW-shift and the Regge limit  can indeed be made precise, it is then possible to adopt much of Regge technology to the BCFW-deformation and also to the study of color-ordered gluon amplitudes beyond the tree-level.   One can also envisage developing more elaborate high energy limits, such as the ``multi-Regge limits", involving multiple-BCFW shifts~\cite{DelDuca:1995zy,Brower:2008nm,Brower:2008ia,Bartels:2008ce,Bartels:2008sc,Henn:2010ir,Henn:2010bk}.

The identification of a BCFW deformation with a Regge limit might appear problematic at first.    In a BCFW standard treatment, helicity configurations are treated asymmetrically. For instance, for four-point amplitudes, with shift in  adjacent  momenta  with helicities  $(h_1,h_2)$,  the asymptotic behavior at large $|z|$ for various configurations are
\be
A_{(-+)}(z) \sim A_{(++)}(z) \sim A_{(--)}(z)  \sim z^{-1}\; , \quad\quad  A_{(+-)}(z) \sim z^3\; .
\label{eq:large-z}
\ee
(See summary   in Table-\ref{tab:large-z}.)   In contrast, 
the  hallmark of Regge behavior is {\it factorization}, which leads to a {\it symmetric} treatment.   It is well known that string amplitudes  exhibit Regge behavior.  It is also well known that, open-string amplitudes involving external (massless) spin-$1$ particles, i.e., gluons,  reduce to YM tree graphs in the zero-slope limit, ($\alpha'\rightarrow 0$.) For example,  a color-ordered 4-gluon flat-space string amplitudes has a universal large $s$ limit at fixed $t$,
\be
A_{string}(s,t) \simeq \gamma_{12}(t) \frac{(-\alpha' s)^{\alpha(t)}}{\sin{\pi \alpha(t)}} \gamma_{34} (t)  +\cdots,
\label{eq:4string}
\ee
where $\alpha(t)$ is the leading $t$-channel Regge trajectory. In flat-space, $\alpha(t)=1+\alpha' t$, which reflects the spin-1 nature of gluon exchange at $t=0$.    Because of the factorization property,  the associated Regge residue, $\gamma_{ij}$,  depends only on the helicities $(h_i,h_j)$, as well as on the momentum transfer invariant, $t$. In such a treatment, helicity configurations $(+,-)$ and $(-,+)$ are on an equal footing, as summarized under column 4 and column 5 in Table-\ref{tab:large-z}.

 In this paper, we first show how the  large-$z$ asymmetry in helicity configurations can be removed in a {\it modified} BCFW treatment where standard BCFW shifted amplitudes are modified by ``dressing factors", as listed in column 6 in Table-\ref{tab:large-z}. In such a modified treatment, the large-$z$ behavior is in agreement with Regge expectation, (column 4 and 5 in Table-\ref{tab:large-z}).  These dressing factors, $\alpha(z)$ and $\beta(z)$, emerge by requiring polarization vectors remain orthonormal under BCFW deformation.

Encouraged by this observation,  we next discuss the possible relation between the large-$z$ behavior for gauge amplitudes  to the existence of ``super-convergence" relations, which had played an important role for conceptual advances leading to the formulation of early string theories.  For 4-point scattering, we show that, by identifying kinematic ``zeros",  convergent amplitudes satisfying unsubtracted dispersion relations  can be constructed.  This d�pends crucially on amplitudes involving spins, a point emphasized in ~\cite{ArkaniHamed:2008gz}.

Finally, we  examine more closely the connection of the BCFW shift to the standard Regge limit.   It is well-understood that a Regge limit can be characterized by a Lorentz boost along a light-cone direction.  We demonstrate that the BCFW parameter $z$ can indeed be related to the rapidity of a Regge boost, $\eta$. This identification can be made precise by working with $O(2,2)$ signature, and we find
\be
\eta=\log (1-i\; z)\; .
\ee

In Sec. \ref{sec:BCFW}, we first provide a quick review of the standard BCFW shift and discuss its possible  connection to the Regge limit. The apparent disagreement in the large-$z$ behavior is further illustrated by examining the 4-point Parke-Taylor amplitudes.  In Sec. \ref{sec:DressingFactor}, we show how the standard BCFW shift can be modified, Eq. (\ref{eq:dressing}),  so as to be in agreement with the Regge expectations. This is accomplished by the introduction of ``dressing factors", which are given by Eq. (\ref{eq:alphabeta}). We also provide in Sec. \ref{sec:String} a more detailed discussion on the Regge expectation based on flat-space string amplitudes. In Sec. \ref{sec:superconvergence}, we briefly review the concept of ``super-convergence" for 4-point amplitudes and demonstrate how more convergent ``reduced amplitudes" can be constructed for all helicity configurations. In Sec. \ref{sec:Regge}, a more precise connection between BCFW shift and Regge is made. This can best be done by continuing to $O(2,2)$ signature.  We end with a brief summary and discussion in Sec. \ref{sec:conclusion}.

\vskip 15pt

\begin{table}[h]
\begin{center}
\begin{tabular*}{152mm}{@{\extracolsep\fill}||c||c|c||c|c||c||}
\hline
\hline
 & \multicolumn{2}{c||}{BCFW Shift}   &
 \multicolumn{2}{c||} {Regge Expectation}       &   Dressing Factor   \\
 \hline
$\;\;\; (h_1,h_j)\;\;\;$  & 
 \multicolumn{1}{c|}{adjacent}
     &   {non-adjacent}           &    \multicolumn{1}{c|}{adjacent}
     &   {non-adjacent}     &    \\

\hline\hline
$\;\;\; (+,-)\;\;\;$ &{ $z^3$    }
	   &  {$z^2$   }          &      $z$ &  $z^0$  &   $\alpha(z)^{-2}\beta(z)^{-2}\sim z^{-2} $ \\
\hline
$\;\;\; (-,+)\;\;\;$   &   $z^{-1}$                                      
     &  {$z^{-2}$   }       &         $z$ &  $z^0$  &   $\alpha(z)^2\beta(z)^{2} \sim z^2$ \\
     \hline
   $\;\;\; (+,+)\;\;\;$      &     $z^{-1} $              
      & $z^{-2}$                  &      $z^{-1}$ &  $z^{-2}$  & $\alpha(z)^{-2}\beta(z)^{2}\sim  1 $  \\
\hline
   $\;\;\; (-,-)\;\;\;$      &   $z^{-1}$              
      & $z^{-2}$                &    $z^{-1}$ &  $z^{-2}$& $\alpha(z)^2\beta(z)^{-2} \sim 1 $  \\
      \hline\hline
\end{tabular*}
\caption{Large-$z$ behavior for various helicity configurations associated with the pair of shifted momenta. Here follow the all out-going (4-momentum and helicity) convention.}
\label{tab:large-z}
\end{center}
\end{table}

\section{BCFW Shift and Regge Limit -- Preliminary:}
\label{sec:BCFW}
The standard BCFW shift  begins by focussing on a specific pair of momenta, $(p_i,p_j)
$, for a color-ordered tree-graph, 
\be
p_i \rightarrow  \hat p_i(z)=p_i+z q\; , \quad\quad
p_j \rightarrow  \hat p_j(z) =p_j-zq\; , 
\label{eq:BCFW0}
\ee
where $q$ is null, and orthogonal to both $p_i$ and $p_j$, i.e., $q^2=q\cdot p_i=q\cdot p_j=0$.  Without loss of generality, we shall always choose $p_i=p_1$. In a spinor representation, the orthogonality condition  can best be realized  by choosing $q=\lambda(p_1)\bar\lambda(p_j)$, where we have adopted the standard spinor notation, $p=\lambda(p)\bar \lambda(p)$, for $p_1$ and $p_j$.  To be precise, the shift is implemented by choosing $\lambda(q)=\lambda(p_1)$ and $\bar\lambda(q)=\bar\lambda(p_j)$ so that
\bea
\lambda(\hat p_1)&=&\lambda'\equiv \lambda_1 \; , \quad \quad\quad\quad   \bar \lambda(\hat p_1)= \bar\lambda'_1\equiv \bar \lambda_1 + z \bar \lambda_j\; ,\nn
\lambda(\hat p_j)&=&  \lambda'_j\equiv \lambda_j - z  \lambda_1\; ,  \quad\; \bar \lambda(\hat p_j)=\bar\lambda'_j\equiv \bar \lambda_j\; ,
\label{eq:BCFW}
\eea
where we have simplified the expression by writing $\lambda_k=\lambda(p_k)$ and $\bar \lambda_k=\bar\lambda(p_k)$.  

We shall first examine in this Section the BCFW shift purely from the perspective of invariants. It is useful to   begin by first consider each color-ordered helicity amplitude in the physical region as a function of Mandelstam invariants, e.g., $A(s,t)$, for $n=4$,  with  $t=(p_1+p_2)^2$, $s=(p_2+p_3)^2$.  The physical region for the s-channel corresponds to $s>-t>0 $.  (The third Mandelstam invariant, $u=(p_3+p_1)^2$, is constrained  by $s+t+u =0$, which follows from momentum conservation. For the s-channel physical region, we also have $u<0$.) Applying BCFW deformation for the pair $(p_1,p_2)$, it follows that $s$ increases linearly with $z$,
\be
s(z)\equiv 2 p_2(z)\cdot p_3 = s(0) + b \; z
\label{eq:s-z}
\ee
where $s(0)= 2p_2\cdot p_3$ and $b=- 2 p_3\cdot q$, with   $t$ unshifted, $
t\equiv 2 p_1(z)\cdot p_2(z) =2p_1\cdot p_2$.
Therefore, the limit of $z$ large appears as the same  limit  for the amplitude $A(s,t)$ for large $s$, with $t$ fixed. That is, under a BCFW deformation with $t$ fixed, large-$z$ limit is the same as the large-$s$ limit,
\be
\lim_{s(z)\rightarrow \infty}  A(s(z), t)\; .
\label{eq:tRegge}
\ee
Kinematically, this precisely  corresponds to the Regge limit.  Alternatively, we can consider shifting a non-adjacent  pair. For $n=4$, without loss of generosity, we choose to  deform the pair  $(p_1, p_3)$. It follows  that  $s(z)=2p_2\cdot p_3(z)$, again grows with $z$ linearly,  
with  $u\equiv 2 p_1\cdot p_3$ fixed. Therefore, the large-$z$ limit for this non-adjacent deformation  corresponds to the Regge limit at fixed-$u$, 
\be
\lim_{s(z)\rightarrow \infty}  A(s(z), u)\; .
\label{eq:uRegge}
\ee
Above construct  can be generalized to the case where $n>4$~\cite{Brower:2008ia,Detar:1971gn,Detar:1971dj,Brower:1974yv}.

  To check this assertion, we can consider the case of $n=4$ and $n=5$ where all tree graphs can simply be expressed by the Parke-Taylor formula for MHV amplitudes. For simplicity, we consider here for $n=4$ in order to establish the Regge expectation. 
Let's begin by expressing  the Parke-Taylor Amplitudes in terms of invariants. For four-point amplitudes with negative helicities for the momentum pair $(k,\ell)$, i.e, $h_k=h_\ell=-$, up to coupling constant dependence which we drop, the amplitudes are given by 
$
A(1,2,3,4)=\frac{\langle k\ell\rangle^4}{\langle 12\rangle\langle 23\rangle\langle 34\rangle\langle 41\rangle}
$. 
One easily finds that,  for six distinct helicity configurations, 
\bea
A_{(+,-,-,+) }(s,t)&=& A_{(-,+,+,-)}(s,t)\; =\;- \frac{s}{t} \; , \nn
A_{(+,+,-,-)} (s,t) &=& A_{(-,-,+,+)} (s,t) \; =\; - \frac{t}{s}\; ,  \nn
A_{(+,-,+,-)} (s,t) &=& A_{(-,+,-,+)}(s,t)\;=\; - \frac{u^2}{s t} = - \frac{s}{t} - \frac {t}{s} - 2 \;.
\label{eq:PT-amplitudes}
\eea

Let us try to extract the large-$z$ behavior under  BCFW shift, according to (\ref{eq:tRegge}) and (\ref{eq:uRegge}). For  the adjacent shift involving the pair $(1,2)$, 
this corresponds to $s$ large with $t$ fixed. For the configurations $(+,-,-,+)$, $(+,-,+,-)$, $(-,+,-,+)$, $(-,+,+,-)$, by identifying large $s$ with large $s(z)=s(0)+b z$, one immediately observes that  Eq. (\ref{eq:PT-amplitudes})  leads to the Regge expectation
\bea
A_{(+,-,\pm,\mp)}  (s,t) &=& A_{(-,+,\pm,\mp)} (s,t) \rightarrow  z\; ,\nn
A_{(+,+,-,-)}  (s,t) &=& A_{(-,-,+,+)} (s,t) \rightarrow  z^{-1}\; ,
\label{eq:PT-regge.1}
\eea
as tabulated in  column 4 of Table-\ref{tab:large-z}.   Next consider the limit appropriate for the BCFW shift with the non-adjacent pair of momenta $(1,3)$. In this limit, $s(z)=-t(z)-u\sim z \rightarrow \infty$, with $u$ fixed. Note that all terms in  Eq. (\ref{eq:PT-amplitudes}) now  contribute. In particular, for $A_{(+,-,+,-)}$ and $A_{(-,+,-,+)}$, the righthand side can be expressed as $u^2/st$. The corresponding large-$z$ behavior for Regge expectation for six configurations are now 
\bea
A_{(+,\pm,-,\mp) }(s,u) = A_{(-,\pm,+,\mp)} (s,u) &  \rightarrow  &z^0\; ,  \nn
A_{(+,-,+ ,-)} (s,u) = A_{(-, +,- ,+)} (s,u)&  \rightarrow & z^{-2} \; ,
\label{eq:PT-regge.2}
\eea
as tabulated in column 5, Table-\ref{tab:large-z}.  Clearly, there is a discrepancy between Eqs. (\ref{eq:PT-regge.1})-(\ref{eq:PT-regge.2}) and  the standard BCFW-results, listed in column 2 of Table-\ref{tab:large-z}.

\section{Modified  BCFW Deformation}
\label{sec:DressingFactor}

We will now examine the BCFW shift more closely in order to resolve the apparent disconnect between its large-$z$ behavior and the Regge expectation, tabulated in Table-\ref{tab:large-z}. Let us first recall some elementary properties   of spinor calculus. Every 4-momentum $p$ can be represented  by a $2\times 2$ matrix
$p_{\alpha\dot{\alpha}}=(p^{\mu}\sigma_{\mu})_{\alpha\dot{\alpha}}$.  If $p$ is a null vector, $det \; p=0$, and this matrix can be written in a `dyadic form, and, in particular, can be 
factorized as a  product of holomorphic and anti-holomorphic two-vectors $\lambda(p)$ and $\bar{\lambda}(p)$,
$
p_{\alpha\dot{\alpha}}=\lambda_{\alpha} \bar \lambda_{\dot{\alpha}},
$
where, in terms of LC components,  
\be
\lambda(p)=(\frac{-p_\perp}{\sqrt {p_+}}, \sqrt {p_+})\; , \quad\quad 
\bar \lambda(p)= (\frac{-\bar p_\perp}{\sqrt {p_+}}, \sqrt {p_+})\; .
\label{eq:spinors}
\ee
Here we  have defined the standard LC coordinates, $p_{\pm}=p_0\pm p_z$, $ p_\perp= p_1+ip_2$, and $\bar p_\perp=p_1-ip_2$. When the $4$-momentum $p$ is real, these two-vectors $\lambda$ and $\bar\lambda$ are related to each
other by complex conjugation, $(\lambda)^{*}=\bar{\lambda}$; for $p$ complex, $\lambda$ and $\bar{\lambda}$ are to be treated as independent.  It is also conventional to represent polarization vectors  through these 2-spinors:
\be
\epsilon_{-}=\frac{\lambda\bar{\mu}}{[\lambda\mu]}\,, \quad\quad  \epsilon_{+}=\frac{\mu\bar{\lambda}}{\left\langle \mu\lambda\right\rangle }\; ,
\label{eq:polarizationspinor}
\ee
where $\mu$ is associated with a reference vector.   We note here  that it is always possible to rescale $\lambda$
and $\bar{\lambda}$, $\lambda\rightarrow t\,\lambda$,
and $\bar{\lambda}\rightarrow\frac{1}{t}\,\bar{\lambda}$, without
affecting the value of bispinor $p_{\alpha\dot{\alpha}}=\lambda\bar{\lambda}$,
but conjugate relation is spoiled under this re-scaling.   Under such a re-scaling polarization vectors scale oppositely, $\epsilon_{-}\sim t^2$ and $\epsilon_{+}\sim t^{-2}$. 
This choice,  (\ref{eq:spinors}), assures these polarization vectors are properly normalized in the physical region, a fact which will become important shortly.\footnote{ The $2$-vectors
have branch cut in $p_+$ starting from $p_+=0$. However, note that $p_{\alpha\dot{\alpha}}=\lambda\bar{\lambda}$
is free from such branch cut, and so are the polarization vectors defined
through these $2$-vectors,
$
\epsilon_{\mu}^{-}=\frac{\lambda\bar{\mu}}{[\lambda\mu]}\sim(\sqrt{p_{+}})^{2},\,\epsilon_{\mu}^{+}=\frac{\mu\bar{\lambda}}{\left\langle \mu\lambda\right\rangle }\sim(\sqrt{p_{+}})^{-2}.
$
}

In deriving the  on-shell BCFW  relation, one begins by first performing a shift on a pair of external momenta,
$(i,j)$. As mentioned earlier, we shall adopt the convention where $i$ is always chosen to be the first leg, i.e., $i=1$, while allowing $j$ to be between $[2,n-1]$. The shift is defined  by Eq. (\ref{eq:BCFW}),
with the constraints $q^2=q\cdot p_{1}=q\cdot p_{j}=0$.   The standard approach is to adopt the convention $\lambda(q)=\lambda(p_1)$ and $\bar \lambda(q)=\bar\lambda(p_j)$, so that (\ref{eq:BCFW0}) follows.  We shall refer to  this choice as the standard BCFW shift, 
\be
\lambda^{BCFW}(\hat p_1)= \lambda'_1\; , \quad \bar\lambda^{BCFW}_1(\hat p_1) = \bar\lambda'_1\; ;\quad\quad
\lambda^{BCFW}(\hat p_j)= \lambda'_j\; , \quad  \bar\lambda^{BCFW}(\hat p_j)= \bar\lambda'_j  \; ,
\label{eq:BCFWspinors}
\ee
with spinors $\lambda'_1, \bar\lambda'_1, \lambda'_j, \bar \lambda'_j$ given by (\ref{eq:BCFW}).   However, as we show below,  a more general treatment is possible.

\subsection{Modified BCFW Shift and Dressing Factors}
Since we are working with null vectors, following (\ref{eq:spinors}), it is easy to relate $\lambda(\hat p_1)$ and $\bar \lambda(\hat p_1)$ to 2-spinors for $q$ and $p_1$, 
\begin{eqnarray}
\lambda  (\hat p_1)
 & = & \frac{\sqrt{p_{1+}}}{\sqrt{p_{1+}+z\, q_{+}}}\,\left[\lambda_1+z\,\frac{\sqrt{p_{1+}}}{\sqrt{q_{+}}}\,\lambda(q)\right], \nn
\bar{\lambda}(\hat p_1)  & = & \frac{\sqrt{p_{1+}}}{\sqrt{p_{1+}+z\, q_{+}}}\left[\bar{\lambda}_1+z\,\frac{\sqrt{q_{+}}}{\sqrt{p_{1+}}}\,\bar{\lambda}(q)\right],\label{eq:lambda-1}
\end{eqnarray}
and similarly, to relate $\lambda(\hat p_j)$ and $\bar \lambda(\hat p_j)$ to 2-spinors for $q$ and $p_j$,  
\begin{eqnarray}
\lambda  (\hat p_j)
 & = & \frac{\sqrt{p_{j+}}}{\sqrt{p_{j+}+z\, q_{+}}}\,\left[\lambda_j-z\,\frac{\sqrt{p_{j+}}}{\sqrt{q_{+}}}\,\lambda(q)\right], \nn
\bar{\lambda}(\hat p_j)  & = & \frac{\sqrt{p_{j+}}}{\sqrt{p_{j+}+z\, q_{+}}}\left[\bar{\lambda}_j-z\,\frac{\sqrt{q_{+}}}{\sqrt{p_{j+}}}\,\bar{\lambda}(q)\right].    \label{eq:lambda-2}
\end{eqnarray}

Let us turn next to the constraints. In light-cone coordinates the inner product $p\cdot q$ can be expressed
as $p\cdot q=p_{-}q_{+}+p_{+}q_{-}-p_{\perp}\bar{q}_{\perp}-\bar{p}_{\perp}q_{\perp}=\frac{1}{p_{+}q_{+}}(p_{+}q_{\perp}-p_{\perp}q_{+})(p_{+}\bar{q}_{\perp}-\bar{p}_{\perp}q_{+})$
so that the condition $p\cdot q=0$ can be satisfied either because
the holomorphic part is zero, $p_{+}q_{\perp}-p_{\perp}q_{+}=0$,
or because the anti-holomorphic part is zero $p_{+}\bar{q}_{\perp}-\bar{p}_{\perp}q_{+}=0$.
It is customary to adopt 
\be
q_{\perp}=\frac{q_{+}}{p_{1+}} p_{1\perp}\; , \quad\quad 
 \bar{q}_{\perp}=\frac{q_{+}}{p_{j+}}\bar{p}_{j\perp} \; ,
 \label{eq:q-perp}
\ee
so that both constraints $q\cdot p_{1}=0$ and $q\cdot p_{j}=0$ are satisfied.  This also implies that
\be
\lambda(q)=\frac{\sqrt{q_{+}}}{\sqrt{p_{1+}}}\,\lambda_1\; , \quad \quad \bar{\lambda}(q)=\frac{\sqrt{q_{+}}}{\sqrt{p_{j+}}}\,\bar{\lambda}_j\; .
\label{eq:q-bispinor}
\ee
Finally,  the holomorphic and anti-holomorphic conditions above does not constrain the magnitude for 
$q_{+}$;  we are thus allowed
to use the remaining degree of freedom to choose $\frac{q_{+}}{\sqrt{p_{1+}}}\,\frac{\sqrt{p_{j+}}}{p_{j+}}=1$. This amounts to fixing the normalization so that 
$
q_{\alpha\dot{\alpha}}= \lambda_{\alpha} (\hat p_1)  \bar \lambda^T_{ \dot{\alpha}}(\hat p_j)
$
With this normalization, we also achieve the following simplification
\begin{eqnarray}
\lambda  (\hat p_1)&=& \alpha(z)\; \lambda'_1, \quad\quad 
\bar{\lambda}(\hat p_1)  =\alpha(z)^{-1}\, \bar{\lambda}'_1,\label{eq:bcfw5}\\
\lambda  (\hat p_j)&= &\beta(z)^{-1}\,\lambda'_{j},\quad
\bar \lambda  (\hat p_j)  =\beta(z)\,\bar{\lambda}'_{j},\label{eq:bcfw8}
\end{eqnarray}
with ``dressing factors"
\be
\alpha(z)= \frac{\sqrt{p_{1+}+z\, q_{+}}}{\sqrt{p_{1+}}}=\frac{\left\langle \xi,\lambda_{p_{1}+zq}\right\rangle }{\left\langle \xi,\lambda_{1}\right\rangle } ,\quad\quad
\beta(z) = \frac{\sqrt{p_{j+}-z\, q_{+}}}{\sqrt{p_{j+}}}=\frac{[\xi,\lambda_{p_{j}-zq}]}{[\xi,\lambda_{j}]} ,
\label{eq:alphabeta}
\ee
where a reference spinor  is  $\xi=\left(\begin{array}{c}
1\\
0
\end{array}\right)$. 
We shall refer to (\ref{eq:bcfw5}) and (\ref{eq:bcfw8}) as modified BCFW shift. Both $\alpha(z)$ and $\beta(z)$ behave as $z^{1/2}$ as $z\rightarrow\infty$.  Note that the difference between the pair $(\lambda  (\hat p_1),\bar{\lambda}(\hat p_1)) $ and $(\lambda'_1, \bar{\lambda}'_1)$  is simply a rescaling by a factor of $\alpha(z)$. Similarly, $(\lambda  (\hat p_j),\bar{\lambda}(\hat p_j)) $ and $(\lambda'_j, \bar{\lambda}'_j)$ are related by a rescaling by $\beta(z)$. We emphasize  that the holomorphic and anti-holomorphic factorization has been maintained   for both $\hat p_{1,\alpha\dot{\alpha}}$ and $\hat p_{j,\alpha\dot{\alpha}}$.

\subsection{Modified BCFW Amplitudes}

It is now clear that the conventional BCFW shift, (\ref{eq:BCFW}),   is  related to the modified shift described
in  (\ref{eq:bcfw5}) and (\ref{eq:bcfw8}) by $z$ dependent
rescalings.
As mentioned earlier, the continuation described by (\ref{eq:bcfw5})
and (\ref{eq:bcfw8}) preserves conjugate relations in the physical region, and therefore
defines orthonormal polarizations, (\ref{eq:polarizationspinor}). To be more precise, the normalization condition is fixed since
the momentum dependence of a
physical polarization is strictly fixed by the condition that it is
a representation of the Lorentz group (up to a longitudinal gauge
choice). At momentum $k=(1,0,0,1)$ the polarizations are solved from
little group eigenvalue equations as $\epsilon^{\pm}(k)=(0,1,\pm i,0)/\sqrt{2}+\alpha k^{\mu}$.
Since polarizations at generic momentum is related to $\epsilon^{\pm}(k)$
by a boost, we have
\begin{eqnarray}
\left(\epsilon^{+}(p)\right)^{*}\cdot\epsilon^{+}(p) & = & \eta_{\mu\nu}\Lambda^{\mu}{}_{\sigma}\Lambda^{\nu}{}_{\rho}\left(\epsilon^{+\sigma}(k)\right)^{*}\epsilon^{+\rho}(k)\label{eq:norm}\nn
 & = & \eta_{\sigma\rho}\left(\epsilon^{+\sigma}(k)\right)^{*}\cdot\epsilon^{+\rho}(k)\nn
 & = & -1.
\end{eqnarray}
The BCFW shifted $2$-vectors, on the other hand,  pick up additional factors of $\alpha$ and
$\beta$, leading to 
\begin{eqnarray}
\epsilon_{1}^{BCFW,+}=\alpha(z)^{2}\epsilon_{1}^{+}, &  & \epsilon_{1}^{BCFW,-}=\alpha(z)^{-2}\epsilon_{1}^{-},\nn
\epsilon_{j}^{BCFW,+}=\beta(z)^{-2}\epsilon_{j}^{+}, &  & \epsilon_{j}^{BCFW,-}=\beta(z)^{2}\epsilon_{j}^{-}.
\end{eqnarray}
Note that $\alpha$ and $\beta$ carry $p_{+}$ and $q_{+}$ dependence in
addition to factors of $z$.  Finally, it follows that physical amplitudes must be ``dressed" by additional factors
\be
A_{physical}(\hat 1 \dots, \hat j\dots) = \alpha(z)^{-2h_1} \beta(z)^{2h_j} A_{BCFW}(\hat{1}\dots\hat{j}\dots)\; .
\label{eq:dressing}
\ee
Since $\alpha(z)$ and $\beta(z)$ behave as $z^{1/2}$ as $z\rightarrow\infty$, it follows that, under this modified BCFW  deformation, amplitudes are modified by these ``dressing factors", $\alpha^{-2h_1} \beta^{2h_j}$, which are listed in the last column of Table-\ref{tab:large-z}.   Because of these dressing factors, the large-$z$ behavior is now symmetric, and they also agree with the Regge expectation indicated in Table-\ref{tab:large-z}.  To be more explicit, we illustrate in Appendix A how conventional BCFW shift gets modified by these dressing factors for the 4-point MHV Parke-Taylor amplitudes.

\section{Regge Expectation from String Amplitudes:}  
\label{sec:String}

Let us next take a closer look at the the string amplitudes, Eq. (\ref{eq:4string}).  As stressed earlier, the hallmark of Regge behavior is {\it factorization}, which can best be illustrated by the use of the so-called ``BPST vertex operator", first introduced for closed string scattering in Ref. \cite{Brower:2006ea}. Extending to open-strings, in exact analogue to Eq. (3.20) of \cite{Brower:2006ea}, a general color-ordered open-string amplitude takes on the following factorized form in the Regge limit,
\bea
\langle V_1V_2\cdots V_n\rangle &\simeq&   \langle {\cal W}_{R}{\cal V}^-_A\rangle \;  \Pi(\alpha' t) \; \langle  {\cal V}^+_A {\cal W}_{L}\rangle   \nn
&\simeq&   \langle {\cal W}_{R0}{\cal V}^-_A\rangle \;  \Big(\Pi(\alpha' t) \left(-\alpha' s\right)^{1+\alpha' t }\Big ) \; \langle  {\cal V}^+_A {\cal W}_{L0}\rangle\;,
\label{eq:openstringRegge}
\eea
where we have grouped vertex operators $\{V_i\}$ into a left-moving group and right-moving group, $ {\cal W}_{R}$ and $ {\cal W}_{L}$. In the second line, $ {\cal W}_{R0}$ and $ {\cal W}_{L0}$ are evaluated in two respective rest frames related by a large Lorentz boost $\eta$ along $+z$ direction, with $\eta\simeq \log  \alpha' s$.  
Eq. (\ref{eq:openstringRegge}) can be interpreted as the product of the propagator for the gluon Regge trajectory, $\alpha(t) = 1+ \alpha' t$, of the form $ \Pi(\alpha' t) \left(-\alpha' s\right)^{1+\alpha' t }$,     with    $\Pi(\alpha' t) \sim \Gamma(-\alpha(t))$, times the couplings  of the gluon-trajectory to the two sets of vertex operators    $ {\cal W}_{R0}$ and $ {\cal W}_{L0}$, through the gluon-trajectory vertex operators
  \be
{\cal V}^{\pm}_A=(\partial X^{\pm}/\alpha')^{\alpha(t)/2} e^{\mp \, i kX}\; .
\ee
This compact representation, Eq. (\ref{eq:openstringRegge}), is particularly useful in exhibiting the factorization property of Regge behavior~\footnote{The use of these BPST vertex operators have also been explored in a slightly different but related context in \cite{Cheung:2010vn}. Other related studies can also be found in  \cite{Fotopoulos:2010cm,Fotopoulos:2010jz,Garousi:2010er,Velni:2012sv,Boels:2008fc}.}. Here ${\cal W}_{R0}$ and $ {\cal W}_{L0}$ can be any string states, including multi-particle states involving both ground states and high spin excitations, with their couplings to the gluon trajectory evaluated in a convenient rest frame. This will be done in a separate publication~\cite{Fu:2013xba}.

For a color-ordered 4-gluon string amplitude  at large $s$, ${\cal W}_{R0}$ and $ {\cal W}_{L0}$ each consists of an incoming and outgoing gluons. The leading behavior  can easily be evaluated, leading to Eq. (\ref{eq:4string}), i.e., it is given by  exchanging the leading $t$-channel Regge pole,  which we repeat here,
\be
A_{string}(s,t) \simeq \gamma_{12}(t) \frac{(-\alpha' s)^{\alpha(t)}}{\sin{\pi \alpha(t)}} \gamma_{34}(t)  +\cdots
\label{eq:4string2}
\ee
where we have replaced the propagator $\Pi(\alpha' t)$ by a more conventional factor $1/\sin{\pi \alpha(t)}$. Note that, from the perspective of a YM theory, the leading ``effective spin" of the exchange is given by a linear trajectory, $\alpha(t) = 1+ \alpha' t$. At $t=0$, this leads to spin-1 exchange, appropriate for the exchange of a gluon, obeying $t$-channel factorization. The associated Regge residue factorizes into a product, with $\gamma_{ij}$  depending on the helicity $(h_i,h_j)$, as well as on the momentum transfer invariant, $t$. For $\gamma_{12}$, the helicity configurations $(+,-)$ and $(-,+)$ are treated on an equal-footing, and, up to a possible phase which is conventional, $\gamma_{(+,-)}=\gamma_{(-,+)}$. This symmetry also holds for $(+,+)$ and $(-,-)$ configurations. 
 
 It is useful to point out one significant difference between string amplitudes and multi-gluon amplitudes. Because of factorization, amplitudes with only one negative helicity or only one positive helicity are both non-zero, i.e., $A_{(-+,+,+)} \neq 0$, $A_{(+,+, +, -)}\neq 0$, etc.  In fact, it is easy to demonstrate, e.g., using the vertex operator technique introduced in \cite{Brower:2006ea},  that the Regge residue in (\ref{eq:4string2}) is given by 
 \be
 \gamma_{ij}\sim \epsilon_{h_i}\cdot  \epsilon_{h_j} - 2\alpha' \epsilon_{h_i} \cdot p_j \; \epsilon_{h_j} \cdot p_i \; .
 \label{eq:reggeoncoupling}
 \ee
 Here $\epsilon_{h_k}$ is the polarization vector~\footnote{Since Regge limit singles out a LC direction, we choose conveniently $p_k$ to move along  the $z$-axis, with either large $p_k^+$ or $p_k^-$, and $p_k\cdot \epsilon_{h_k}=0$. Each polarization vector also satisfies an auxiliary orthogonality condition. For particle with large $+z$ 3-momentum, we choose a reference vector $\tilde p$ where $\tilde p^+=0$ so that $\tilde p_k \cdot \epsilon_{h_k}=0$. For particle with large $p^-$, we choose a reference vector where $\tilde p^-=0$.  Equivalently, one can make use of the spinor representation, (\ref{eq:polarizationspinor}).}  for the kth particle with momentum $p_k$ and helicity $h_k$,  (adopting all-outgoing convention).  It follows that $ \epsilon_{h_j}\cdot \epsilon_{h_j}  \sim \delta_{h_j,-h_k}$, which can be seen more directly  by using the spinor representation, (\ref{eq:polarizationspinor}). Furthermore, in the forward scattering limit, the amplitude is ``$s$-channel helicity preserving", and $\gamma_{(+,+)}=\gamma_{(-,-)}=0$  at $t=0$. From  (\ref{eq:reggeoncoupling}), one finds these couplings vanish as $ \alpha(t) -1$, i.e.,
 \be
 \gamma_{(+,+)}(t) =\gamma_{(-,-)}(t) \sim \alpha' \; t\; .
 \ee
   This vanishing behavior can be understood kinematically, and there are no other asymmetry in the Regge limit of large-$s$. In particular, in (\ref{eq:4string}), the dominant power of $s$, $\alpha(t)$, is universal.

It is also well-known that one can recover  the color-ordered gauge amplitude from the corresponding string amplitude by  taking the ``zero-slope" limit. 
  In the limit $\alpha'\rightarrow 0$,  one finds $\alpha(t)\rightarrow 1$,  $\sin {\pi \alpha(t)} \rightarrow -\pi \alpha' t$, and
\be
A_{string}(s,t) \rightarrow  \widetilde \gamma_{12}\widetilde \gamma_{34}  \left(\frac{ s}{ t} \right)  +\cdots  
\label{eq:4gluon}
\ee
  It is  important to note that, in this limit,  the coupling  $\widetilde \gamma$ is  no longer a function of $t$, since there are no other dimensionful quantity left to serve as a scale. Since the amplitude is  helicity preserving in the forward limit, it follows that $\tilde \gamma$ remains helicity preserving for all $t$, i.e., $\widetilde \gamma$ vanishes for spin-flip configurations, 
\be
\widetilde \gamma_{++}=\widetilde \gamma_{--}=0\; .
\ee
This fact is now consistent with the  MHV rule for gluon helicity amplitudes, i.e., amplitudes with helicities all positive (negative)  and all but one positive (negative) vanish identically. It follows that the the leading $t$-channel Regge exchange  contributes only  to  four helicity configurations:  $(+,-,-,+)$, $(+,-,+,-)$, $(-,+,-,+)$, $(-,+,+,-)$, with two equal constant couplings $\widetilde\gamma_{+-}$ and $\widetilde\gamma_{+-}$, and
\bea
A_{(+,-,\pm, \mp)} (s,t) = A_{(-,+,\mp,\pm) }(s,t) &\sim  &   \frac{s}{t}  +\cdots
\label{eq:t-exchange-s}
\eea
This linear dependence on $s$ reflects the spin-1 nature of the $t$-channel exchange. 
Turning next to the BCFW shift at large $z$. If we are allowed to identify $s(z)$ with $z$, i.e., (\ref{eq:tRegge}), we then have
\bea
A_{(+,-,\pm, \mp)} (s,t) =   A_{(-,+,\mp,\pm) }(s,t)  &\rightarrow  &z  \; .
\label{eq:t-exchange-z}
\eea
This symmetric dependence on helicity configurations is in agreement with that listed for Regge expectation, under first two entries of column 4 in Table-\ref{tab:large-z}. The large-$z$ behavior for  $A_{(+,+,-,-)}(s,t)$ and $A_{(-,-,+,+)} (s,t)$, the remaining entries, can only be determined if we keep track of the next-to-leading order contributions to Eq. (\ref{eq:4string2}).

Turning next to  the fixed $u$ Regge limit, again with $s(z)\sim z$ but $u$ fixed, i.e., the BCFW deformation for a non-adjacent pair, $(p_1,p_3)$.  It should be emphasized that, for the color ordering we are considering, the string amplitudes $A_{string}(1,2,3,4)$ contains only $s$- and $t$-singularities. It therefore does not contain $u$-channel pole exchange contribution, and the large-$z$ must be determined by keeping both $t$-channel and $s$-channel pole contributions. In fact, with $\alpha'\neq 0$, for such a Regge exchange, the associated residue typically vanishes exponentially, leading to exponentially small contribution.  However,  the limit $\alpha'=0$ changes the behavior dramatically, leading to a power behavior. From Eq. (\ref{eq:4gluon}),  the ratio $s/t$ goes to a constant, in the limit with $u$ fixed.  This constant behavior accounts for the first two entries of column 5 in Table-\ref{tab:large-z}. 

It should be stressed that, in this fixed-u limit, the contribution from the lower order terms must be taken into account. From the Parke-Taylor formula, we note the importance of the contribution from the contact term, which cancels the contributions from the direct-channel $s$- and $t$-pole, leading to the $z^{-2}$ behavior, for the last two entries in column 5.  Since fixed-u limit, strictly speaking, cannot be associated with u-channel exchanges, one should refrain  from interpreting its large-z behavior as due to Regge exchange.

\section{Kinematic-Singularities and Super Convergence}
\label{sec:superconvergence}
It is now clear that a deeper understanding on the asymptotic behavior of analytic amplitudes can play an important role in expanding the BCFW recursive approach to the study of  the dynamics of gauge theories such as QCD. Regge asymptotics not only provides a more systematic control on the convergence of dispersion integrals, it has historically also served as the starting point in relating ``cross-channel exchanges" to direct channel singularities~\footnote{Historically, this is referred to as ``Dolen-Horn-Schmidt" duality~\cite{Dolen:1967zz,Dolen:1967jr}.}. Indeed, the recognition of  ``s-t duality" directly led  to the original  dual-model, which, in turn, led to the birth of the string theory.  In this context, we recall  the discovery of   ``super-convergence sum rules" \cite{DeAlfaro:1966}, i.e., if an amplitudes, $A(s,t)$, at fixed $t$, vanishes more rapidly than $O(1/s)$, as $|s| \rightarrow \infty$, this leads  to a sum rule for the imaginary part of $A(s,t)$,
\be
\int ds \; Im A(s,t)=0\; .
\ee
Such a ``super-convergence" relation occurs naturally when scattering involves external  spins. We now demonstrate  how such super-convergence-like relations arise for gauge amplitudes. For instance, in spite of the divergent asymptotic behavior for some helicity configurations, e.g., Table-\ref{tab:large-z}, a convergent, reduced amplitude can be designed for all these situations due to existence of ``kinematic zeros".  

Consider scattering in the s-channel center-of-mass frame for the $a+b\rightarrow c+d$. In order to confirm to traditional treatments, we shall initially adopt the usage where $p_a$ and $p_b$ are incoming and $p_c$ and $p_d$ are outgoing, with all zeroth components positive. Denote helicity amplitude by $F_{\lambda_c,\lambda_d;\lambda_a,\lambda_d}(s,t)$, where ``forward scattering" corresponds to $t=(p_c-p_a)^2=0$ and ``backward scattering" corresponds to $u=(p_c-p_a)^2$.  We further assume the spatial momentum $\vec p_a$ is along $+z$ and $\vec p_b$ along the negative $z$-axis.  Therefore, the total angular momentum $J_z$ along the $z$-axis is initially $\lambda\equiv \lambda_a-\lambda_b$. In the forward scattering limit where $t=0$, the final total $z$-component of angular momentum is $\mu= \lambda_c-\lambda_d$. Since $J_z$ is conserved, one finds the amplitude must vanish  if $\lambda\neq -\mu  $, 
\be
F_{\lambda_c,\lambda_d;\lambda_a,\lambda_d}(s,t=0) =0\; .
\ee
 Similarly, for backward scattering where the final z-component $\mu$ is reversed, the amplitude again vanishes if $\lambda\neq -\mu  $,
 \be
 F_{\lambda_c,\lambda_d;\lambda_a,\lambda_d}(s,u=0) =0\; .
 \ee
Therefore, these constraints automatically lead to kinematical zeros on the boundaries of physical region. From the s-channel perspective, kinematic zeroes occur at $t=0$ and $u=0$. By removing these kinematic singularities, one obtains  a reduced amplitude having only ``dynamical singularities"  in the $t-u$ plane, with $s$ fixed~\cite{Trueman:1966zz,Hara:1964zza,Wang:1966zza,Barut:1963zzb}.  For gauge theories and for gravity, interestingly, these are again at $s=0$ and $u=0$.
  
  The above discussion has traditionally been used for scattering of massive particle. It is possible to identify these kinematic zeroes  by a pre-factor
  \be
  \chi_{\mu,\lambda}=\Big ( \sin \frac{\theta}{2} \Big)^{|\lambda -\mu | } \; \Big ( cos \frac{\theta}{2} \Big)^{|\lambda +\mu | }\; .
  \ee
It follows a kinematic-singularity free amplitude can be defined by dividing out this pre-factor
\be
\widetilde F_{\lambda_c,\lambda_d;\lambda_a,\lambda_d}(s,t)= \chi_{\mu,\lambda}^{-1} \; F_{\lambda_c,\lambda_d;\lambda_a,\lambda_d}(s,t)\; .
\ee
At fixed $s$, for this reduced amplitude, singularities in the complex $t$-plane are dynamical. In what follows, we shall adopt this approach, assuming that removal of these kinematic zeros remains meaningful for processes where all particles are massless.

For our present purpose, we are interested in singularities in the complex s-plane for color-ordered amplitudes at fixed t or fixed u. The kinematic factors involved are therefore that associated with the t-channel and u-channel pre-factors respectively. From the t-channel perspective, kinematic zeros are at $s=0$ and $u=0$. For u-channel, they are located at $t=0$ and $s=0$. Returning to our all  outgoing convention, with helicities $(h_1,h_2,h_3,h_4)$ for our color ordering, we have
\bea
 \chi^{(t)}_{\mu_t,\lambda_t}&=& \Big ( \sin \frac{\theta_t}{2} \Big)^{|\lambda_t -\mu_t | } \; \Big ( \cos \frac{\theta_t}{2} \Big)^{|\lambda_t +\mu_t | }  \; ,\nn
  \chi^{(u)}_{\mu_u,\lambda_u}&=&\Big ( \sin \frac{\theta_u}{2} \Big)^{|\lambda_u -\mu_u | } \; \Big ( \cos \frac{\theta_u}{2} \Big)^{|\lambda_u +\mu_u | }
  \eea
where $\lambda_t=h_1-h_2$, $\mu_t=h_3-h_4$, and $\lambda_u=h_3-h_1$, $\mu_u=h_2-h_4$. For scattering involving massive particles, these pre-factors also contain ``threshold singularities". For $ \chi^{(t)}_{\mu_t,\lambda_t}$, it contains  t-channel ``threshold singularities" at $t=(m_a\pm m_c)^2$, $t=(m_d\pm m_b)^2$, and also at $t=0$. At fixed $t$, the t-channel reduced amplitude
\be
\widetilde A^{(t)}_{(h_1,h_2,h_3,h_4)}(s,t)= { \chi^{(t)}_{\mu_t,\lambda_t}}^{-1} \; A_{(h_1,h_2,h_3,h_4)}(s,t)
\ee
contains only dynamical singularities in the complex s-plane.  Similarly, at fixed $u$,
\be
\widetilde A^{(u)}_{(h_1,h_2,h_3,h_4)}(s,u)= { \chi^{(u)}_{\mu_u,\lambda_u}}^{-1} \; A_{(h_1,h_2,h_3,h_4)}(s,t)
\ee
is free from kinematic singularities in s.

  In the massless limit, these kinematic factors simplify, although all singularities are now degenerate. Since there are no other dimensionful scales, they can only depend on ratios invariants. For $ \chi^t_{\mu_t,\lambda_t}$, it  reduces to 
\be
 \chi^{(t)}_{\mu_t,\lambda_t}=\Big ( \frac{s}{t} \Big)^{\frac{|\lambda_t -\mu_t | }{2}}\; \Big ( \frac{u}{t} \Big)^{\frac{|\lambda_t +\mu_t |}{2} }
\ee
and 
\be
\widetilde A^{(t)}_{(h_1,h_2,h_3,h_4)}(s,t)= \Big ( \frac{s}{t} \Big)^{-\frac{|\lambda_t -\mu_t | }{2}}\; \Big ( \frac{u}{t} \Big)^{-\frac{|\lambda_t +\mu_t |}{2} } \; A_{(h_1,h_2,h_3,h_4)}(s,t)\; .
\ee
Similarly, we have
\be
\widetilde A^{(u)}_{(h_1,h_2,h_3,h_4)}(s,u)= \Big ( \frac{t}{u} \Big)^{-\frac{|\lambda_u-\mu_u | }{2}}\; \Big ( \frac{s}{u} \Big)^{-\frac{|\lambda_u +\mu_u |}{2} } \; A_{(h_1,h_2,h_3,h_4)}(s,t)\; .
\ee

In Table-\ref{tab:superconv}, we list  $ \chi^{(t)}_{\mu_t,\lambda_t}$  and  $\chi^{(u)}_{\mu_t,\lambda_t} $ for all 4-point helicity configurations, as well as the reduced amplitudes $\widetilde A^{(t)}_{(h_1,h_2,h_3,h_4)}(s,t)$ and $\widetilde A^{(u)}_{(h_1,h_2,h_3,h_4)}(s,u)$.  There are two interesting observations. First, both $\widetilde A^{(t)}_{(h_1,h_2,h_3,h_4)}(s,t)$  and $\widetilde A^{(u)}_{(h_1,h_2,h_3,h_4)}(s,u)$ satisfy unsubtracted dispersion relations in $s$ at $t$ and $u$ fixed respectively. Second, both $\widetilde A^{(t)}_{(h_1,h_2,h_3,h_4)}(s,t)$ and $\widetilde A^{(u)}_{(h_1,h_2,h_3,h_4)}(s,u)$ are universal, independent of helicity configurations, taking on $-\frac{t}{s}$ and $-\frac{u^2}{s t}$ respectively.
In terms of the BCFW-shift, $z$, we have at large $|z|$, 
\be
\widetilde A^{(t)}(s,t)\rightarrow z^{-1}\; , \quad\quad \widetilde A^{(u)}(s,u)\rightarrow z^{-2}\; . 
\label{eq:reduced}
 \ee

\begin{table}[h]
\begin{center}
\begin{tabular*}{160mm}{@{\extracolsep\fill}||c|c||c|c|c|c||c|c|c|c||}
\hline
\hline
 &  & \multicolumn{4}{c||}{$t$ fixed, $s \rightarrow \infty$}  &
 \multicolumn{4}{c||} {$u$ fixed, $s  \rightarrow \infty$}           \\
 \hline
  $\;\;\; (h_1,h_2,h_3,h_4)\;\;\;$ &$A(s,t)$  & 
 \multicolumn{1}{c|}{$\lambda_t$}
     &  $\mu_t $   &$\chi^{(t)}_{\lambda_t,\mu_t}$ & $ \widetilde  A^{(t)}(s,t)$   &    \multicolumn{1}{c|}{$\lambda_u$}
     &  $ \mu_u $ & $\chi^{(u)}_{\lambda_u,\mu_u}$   &  $\widetilde  A^{(u)}(s,u)$  \\

\hline\hline
$\;\;\; (+,-,-,+)\;\;\;$& -  $\frac{s}{t}$ & $\;\; 2$    
	   &  $-2$  &   $(\frac{s}{t})^2$ & -$\frac{t}{s}\rightarrow z^{-1}$      &      $-2$ &  $-2$&   $(\frac{s}{u})^2$ &  - $\frac{u^2}{st}\rightarrow z^{-2}  $  \\
\hline
$\;\;\; (-,+,+,-)\;\;\;$  &  - $\frac{s}{t}$ &   $-2$                                      
     &  $\;\;2$   &  $(\frac{s}{t})^2$ &-$\frac{t}{s}\rightarrow z^{-1}$    &         $\;\; 2$ &  $\;\; 2$  & $(\frac{s}{u})^2$  &   - $\frac{u^2}{st}\rightarrow z^{-2}  $ \\
     \hline
   $\;\;\; (+,+,-,-)\;\;\;$   &  -$\frac{t}{s}$ &     $\;\;0$              
      & $\;\;0$   & 1 &-$\frac{t}{s}\rightarrow z^{-1}$           &      $-2$ &  $\;\; 2$& $(\frac{t}{u})^2$   &  - $\frac{u^2}{st}\rightarrow z^{-2}  $ \\
\hline
   $\;\;\; (-,-,+,+)\;\;\;$   & -$\frac{t}{s} $ &   $\;\;0$              
      & $\;\;0$ &  1 & -$\frac{t}{s}\rightarrow z^{-1}$          &    $\;\; 2$ &  $-2$&  $(\frac{t}{u})^2$ &  - $\frac{u^2}{st}\rightarrow z^{-2}  $ \\
      \hline
   $\;\;\; (-,+,-,+)\;\;\;$   &  -$\frac{u^2}{st} $ &     $-2$              
      & $-2$ & $(\frac{u}{t})^2$ & -$\frac{t}{s}\rightarrow z^{-1}$              &      $\;\;0$ &  $\;\;0$&  1  &  - $\frac{u^2}{st}\rightarrow z^{-2}  $  \\
\hline
   $\;\;\; (+,-,+,-)\;\;\;$  &  -$\frac{u^2}{st}  $  &   $\;\;2$              
      & $\;\;2$ &  $(\frac{u}{t})^2$   &-$\frac{t}{s}\rightarrow z^{-1}$       &    $\;\;0$ &  $\;\;0$&   1 &  - $\frac{u^2}{st}\rightarrow z^{-2}  $ \\
      \hline\hline
\end{tabular*}
\caption{Kinematic-Singularity-Free Amplitudes}
\label{tab:superconv}
\end{center}
\end{table}

It is worth emphasizing  the intrinsic asymmetry between $\widetilde A^{(t)}(s,t)$ and $\widetilde A^{(u)}(s,u)$.  Due to color-ordering, the original color-ordered amplitudes $A_{(h_1,h_2,h_3,h_4)}(s,t)$ can only have singularities in $s$ and $t$.  At fixed t, $\widetilde A^{(t)}(s,t)$ only has ``right-hand" singularities, i.e., at $s=0$. In contrast, at fixed u, $\widetilde A^{(u)}(s,u)$ can have both left-hand and right-hand singularities, i.e., poles at $s=0$ as well as at $t=0$.  Also due to the s-t symmetry, it follows that $\widetilde A^{(u)}(s,u)$ satisfies a super-convergent sum rule.  

Finally, we  also note that, for graviton scatterings, the distinction between $\widetilde A^{(t)}(s,t)$ and $\widetilde A^{(u)}(s,u)$ disappear and both have both left-hand and right-hand singularities. However, due to spin-2 nature of exchange, we obtain super-convergence where
\be
\widetilde A^{(t)}(s,t)= \widetilde A^{(u)}(s,u)\rightarrow z^{-2}\; .
\label{eq:reduced-graviton}
 \ee
We hope to be able to discuss the  issue of graviton scattering  in a separate treatment.

\section{Regge Limit and $O(2,2)$ Analysis}
\label{sec:Regge}

We now focus on  establishing  a more precise connection between the BCFW shift with the general Regge limit. As discussed earlier, for string theories, Regge behavior  can best be demonstrated  through the used ``vertex operator", which follows from a standard OPE analysis~\cite{Brower:2006ea}.  More generally,  in a traditional treatment of Regge limit, one begins with a partial-wave expansion in the crossed channel of $t>0$ and  $s<0$, and next analytically continue the amplitude to the the direct-channel of $s>0$ and $t<0$ via a Sommefeld-Watson transformation in complex angular momentum, $j$~\cite{Chew,Gribov,DTFT}.  Alternatively, by staying in the physical region of $s>0$ with $t<0$, the Regge limit of large $s$ can be characterized by a Lorentz boost, specified a large rapidity, $\eta$, along a light-cone direction, and, without loss of generality, it is convenient to choose the boost  to be along the positive LC, i.e., the positive $3$-axis.  To be more precise, consider the case of BCFW shift involving adjacent momenta $(p_1,p_2)$. As done earlier for string amplitudes, it is convenient to divide all  momenta into two groups, the right-moving group, consisting of $(p_1,p_2)$, and all the rest, $(p_3,p_4, \cdots)$, in the  left-moving group. Each group is initially specified in a right- and left-Lorentz frame, and  these two frames are connected by the  aforementioned Lorentz boost in the $+z$ direction.  As such, Lorentz invariants made out of momenta within each group will be unaffected by the boost, e.g., $t=(p_1+p_2)^2$, whereas invariants involving momenta from both will increase with the boost, e.g.,    $s=(p_2+p_3)^2\sim e^{\eta} $ for $\eta$ large. 

It is now possible to consider the amplitude as a function over  the Lorentz boost, leading to an expansion in terms of the principal continuous  representation of $O(2,1)$,  labelled by an index  $j$. This representation can essentially be treated as an inverse Mellin transform,
\be
A(s, t) = \int \frac{dj}{2\pi i}  a(j,t) {\cal D}_j( \eta) \sim  \int \frac{dj}{2\pi i} \; \frac{(-s)^j}{\sin\pi j}\; a(j,t) 
\label{eq:Regge}
\ee
 for $\eta$ large where $s\sim e^\eta$. Here we have suppressed all other boost independent invariants except one, the invariant mass squared $t$. Regge exchange comes from the dynamics, which is made possible  by  an analytic continuation to the $t$-channel physical region of $t>0$ and $s<0$,  leading to the identification of $t$-dependent poles in the $j$-plane for the ``partial-wave" amplitude $a(j,t)$\footnote{For simplicity, we have not exhibited various j-dependent factors  in (\ref{eq:Regge}) except for the factor $\sin\pi j$.    The contour in $j$ is to the right of all poles of $a(j,t)$ and to the left of poles $\sin\pi j$ at positive integers. This corresponds to an inverse Sommefeld-Watson transform. This version of Regge hypothesis was first advocated by M. Toller in late 1960's, \cite{Toller:1968zz,Toller:1969vt,Toller:1969gx}}.  As shown in \cite{Brower:2006ea}, for flat-space open-string theory, this identification for the leading Regge trajectory also follows from the on-shell condition: $L_0=1$.  In particular, if a Regge pole exists, the contour above must be deformed to expose this pole, leading to a dominant contribution as exhibited in (\ref{eq:4string2}).

We are now in a position to connect a BCFW shift with taking a Regge limit in the physical region. To facilitate the discussion, we shall first restrict ourselves to a 4-point amplitude, $M(1,2,3,4)$, where we identify invariants $s=(p_1+p_4)^2=(p_2+p_3)^2$, $t=(p_1+p_2)^2=(p_3+p_4)^2$, and $u=(p_1+p_3)^2=(p_2+p_4)^2$, where, due to momentum conservation, $s+t+u=0$.  Let us begin in the $s$-channel physical region, $(s_0,t_0)$ where $s_0>0$, and $t_0<0$. It is useful to consider  $P=\{p_1, p_2, p_3, p_4\}$  collectively as a point in the phase space, subject to on-shell conditions as well as energy-momentum conservation.  Taking  the Regge limit of large $s$ at fixed $t$  traces out a path in the phase space, which  can be parametrized by a one-parameter curve. This path  can be chosen by varying $p_1$ and $p_2$, i.e.,
\be
P(\eta)=\{p_1(\eta), p_2(\eta), p_3, p_4\}\; .
\ee
Along this path, $t(\eta)=t_0$ is fixed,  and $s(\eta)\rightarrow \infty$ as $\eta$ increases, with $s(0)=s_0$. (A corresponding path can also be constructed for the Regge limit with $u$ fixed by varying $p_1$ and $p_3$, e.g., $P(\eta)=\{p_1(\eta), p_2, p_3(\eta), p_4\}
$.) This one-parameter path can be carried out by a Lorentz boost, along the positive LC, as described earlier.  It is also customary to choose the parameter $\eta$ as the rapidity. Once components of momenta for the initial point are given, $P(0)$, it is relatively easy to find the explicit parametrization for $P(\eta)$.  

We next  demonstrate that taking $s$ from $s_0$ to $\infty$ with $t<0$ fixed can also be accomplished by a BCFW shift, with an appropriately chosen null vector $q$, with the path parametrized by the BCFW parameter $z$.   Since this path should lead to the Regge limit, which is characterized by a boost along the positive LC direction,  the BCFW null vector $q$ should point along the same LC direction, i.e., we choose
\be
q^-=q_\perp=\bar q_\perp=0\; ,
\label{eq:LC}
\ee
with $q^+$ a parameter to be specified later\footnote{This condition, (\ref{eq:LC}), is not strictly necessary but it helps to simplify the discussion.}. We further note that, as a physical scattering process, we  shall consider  an initial configuration where $p_1$ and $p_4$ are incoming, thus with negative energy components, and $p_2$ and $p_3$ are outgoing, with positive energy components.  This can be further simplified, without loss of generality,  by choosing a frame where $p=p_1+p_4$ is longitudinal, i.e., lies in the $0-3$ plane, with only $p^{\pm}$ components, 
\be
s_0= p^2=p^+p^-\;.
\ee
We can also choose  $Q=p_2+p_1$ to be  transverse, i.e., in the $1-2$ plane, with $Q^\pm=0$ and 
\be
t_0=Q^2= - Q_\perp \bar Q_\perp\; .
\ee

By imposing the constraints of BCFW construct, $q\cdot p_1=q\cdot p_j=0$ and the massless condition $q^2=0$, we have been led to Eq. (\ref{eq:q-bispinor}).  This condition also  maintains  holomorphic factorization, (\ref{eq:q-perp}). It is easy to convince oneself that, due to these constraints, the Regge constraint of having $q$ pointing in the positive LC direction, (\ref{eq:LC}), cannot be met by having real vectors. On the other hand, this can be achieved by first continuing all gluon momenta to be in the $O(2,2)$ region, i.e., by treating $p_\perp$ and $\bar p_\perp$ as real and  independent~\footnote{A study of amplitudes using $O(2,2)$ continuation has recetnly been carried out in a related context in \cite{Srednyak:2013ylj}.}. There is no obstruction in this continuation while keeping the Mandelstam invariants in the physical region where $s_0>0$, $t_0<0$ and $u_0<0$. 

Continuation to $O(2,2)$ signature can be made more explicit by introducing unit vectors: $\hat e^{\pm} = (1/\sqrt 2) (\hat e_0 \pm \hat e_3) $,  $\hat e_\perp = (1/\sqrt 2) (\hat e_1 + i\; \hat e_2) $ and $\hat {\bar e} _\perp = (1/\sqrt 2) (\hat e_1 - i \; \hat e_2) $, and each 4-vector will be represented by its $O(2,2)$ components:   $p_k= (p_{k+}, p_{k-} ; p_{k\perp} , \bar p_{k\perp})$.  For our initial configuration, we can choose 
\bea
&p_1= (-p_{+}, 0; 0, \bar p_{\perp})\;,  \quad &p_3= (0, p_{-};- p_{\perp},0)\;,  \nn
&p_2=  (p_{+}, 0;  p_{\perp}, 0)\;,  \quad & p_4= (0, -p_{-}; 0, -\bar p_{\perp})\;,
\label{eq:basic-conf}
\eea
with  $p=p_1+p_4=-(p_2+p_3)=(-p_+,-p_-; 0,0)$, $Q=p_1+p_2=-(p_3+p_4)=(0,0;p_\perp, \bar p_\perp)$, and
\be
s_0=p_{+}p_{-} \;, \quad t_0=- p_{\perp} \bar p_{\perp}\; .
\ee
In this frame, a Lorentz boost with rapidity $\eta$ in the positive LC direction acting on $p_1$ and $p_2$, leads to $p_{1+}(\eta) = p_{2+}(\eta)=-e^\eta p_{+}$, and 
\be
 s_0\rightarrow s(\eta) = e^\eta  s_0
\ee
with $t$ fixed.  

Instead of a Lorentz boost, what if we perform a BCFW shift? With the bispinor $q\cdot\sigma$ given by $\lambda_1\bar\lambda_2$, it follows from (\ref{eq:basic-conf}) that $q^+=\sqrt{p_{1+}}\sqrt{p_{2+}} = i p_+$ and $q^-=q_\perp=\bar q_\perp=0$, consistent with (\ref{eq:LC}). The factor of $i$ can be absorbed by the shifting parameter,  $z\rightarrow iz$, so that $q\rightarrow -i\lambda_1\bar\lambda_2$ is real. The shifted vector $\hat p_1$ and $\hat p_2$ now takes on 
\be
\hat p_1=  (- (1-iz)p_{+}, 0; 0, \bar p_{\perp})\; ,\quad \quad
\hat p_2= ((1-iz)p_{+}, 0;  p_{\perp}, 0)\; .
\ee
These components remain real if z is purely imaginary. As expected, $Q=\hat p_1 + \hat p_2$ remains unchanged by the BCFW shift whereas $p=\hat p_1+p_4=(-(1-iz)p_+, -p_; 0,0)$ and
\be
s(z)=p^2=(1-iz)p_+p_-=(1-iz) s_0\; .
\label{eq:sz}
\ee
It follows that a Regge boost has been achieved with $z$ relating to the rapidity by
\be
\eta = \log (1-i\; z) \; .
\ee
This is the key result.  It is also worth noting that the dressing factors $\alpha(z)$ and $\beta(z)$ now take on simple form
\be
\alpha(z)=\beta(z)=\sqrt{1-iz}=\sqrt{s(z)/s_0}\; .
\label{eq:alphabeta2}
\ee
Therefore, these dressing factors can serve  the purpose of providing the appropriate factors of kinematic zeros such as that discussed in Sec. \ref{sec:superconvergence}.  For further illustration, this will be discussed in Appendix A by considering Parke-Taylor amplitudes  under the shift BCFW shift.

\section{Discussion}
\label{sec:conclusion}

In this paper, we have taken a step towards establishing the connection between the BCFW deformation and the more traditional Regge limit.  We begin by   first showing  how the  large-$z$ asymmetry in helicity configurations in a conventional BCFW deformation can be removed in a {\it modified} BCFW treatment involving additional  ``dressing factors", as listed in column 6 in Table-\ref{tab:large-z}. In such a modified treatment, the large-$z$ behavior is in agreement with Regge expectation, (column 4 and 5 in Table-\ref{tab:large-z}).   

Encouraged by this observation,  we next discuss the possible relation between the large-$z$ behavior for gauge amplitudes  to the existence of ``super-convergence" relations, which had played an important role for conceptual advances leading to the formulation of early string theories. Both depend crucially on amplitudes involving spins, a point which has been emphasized in Ref. \cite{ArkaniHamed:2008gz,ArkaniHamed:2008yf}.  For 4-point scattering, we show that convergent amplitudes satisfying UDR  can be constructed. To be precise, we find, from Eq. (\ref{eq:reduced}),  for {\it all} helicity configurations,
\be
\widetilde A^{(t)}(s,t)\rightarrow z^{-1}\; , \quad\quad \widetilde A^{(u)}(s,u)\rightarrow z^{-2}\; . 
 \ee
Finally, we  examine more closely the connection of the BCFW shift to the standard Regge limit.   It is well-understood that a Regge limit can be characterized by a Lorentz boost along a light-cone direction.  We demonstrate that the BCFW parameter $z$ can indeed be related to the rapidity of a Regge boost, $\eta$. This identification can be made precise by working with $O(2,2)$ signature, and we find
\be
\eta=\log (1-i\; z)\; .
\ee
To keep  the discussion simple, we have focused on four point amplitudes where the notion of Regge limit is relatively easy to illustrate, e.g.,  using four-point Veneziano amplitude. The generalization to the case involving higher point functions can be carried out fairly straight forwardly in the case of ``adjacent deformation", which, kinematically, is the nearly identical to a four-point function, as far as the color-ordering is concerned. As for Regge behavior, the separation into two groups of right-moving and left-moving gluons remains the same.  The only added new feature being, in taking the Regge limit, many invariants now become large while their ratios remain fixed. 

For  non-adjacent deformation, many more color-ordering can now be involved as the number of gluons increases. This leads to the so-called ``helicity-pole" limit~\cite{Detar:1971gn,Detar:1971dj,Brower:1974yv,
Brower:2008ia}, which is one of many more elaborate high energy limits, such as the ``multi-Regge limits"~\cite{DelDuca:1995zy,Brower:2008nm,Brower:2008ia,Bartels:2008ce}. It is possible to examine these limits  involving multiple-BCFW shifts, and we will treat these in future publications.

\section{Acknowledgments}

We would like to  thank Bo Feng, Marcus Spradlin and Anastasia Volovich   for
helpful discussions. This work is supported in part by the National Science
Council, 50 billions project of Ministry of Education, National Center for
Theoretical Sciences and S.T. Yau center of NCTU, Taiwan. The work of C.-IT. was also supported in part by the Department of Energy under contract  DE-FG02-
91ER40688, Task-A.

\appendix

\appendix
\def\theequation{A.\arabic{equation}}
\setcounter{equation}{0}

\section{Brief Review of Regge Theory:}
This is a short pedagogical review for those who have not had prior exposure to the basics of Regge theory.  In particular, we discuss only 4-point amplitudes, $A(s,t)$, for the scattering of 4 scalar particles. Regge theory can also be applied to multiparticle scattering,~\cite{Detar:1971gn,Detar:1971dj,Brower:1974yv,
Brower:2008ia}, which we will not deal with here.

Regge theory was used rather successfully to study the high energy behavior of scattering amplitudes in the 1970's. Prior to its phenomenological application, it was recognized in early 60's that in any bootstrap program based on analyticity and unitarity, asymptotic behavior of amplitudes must be determined dynamically.  The work of Regge in potential theories where the corresponding asymptotic behavior (in the unphysical limit of  large momentum transfer)  is controlled  by the bound-state ``spin-energy curves", now referred to as the  Regge trajectories, thus provided the necessary tool in implementing this program. It is interesting to note that  the Veneziano model, the precursor to the modern string theories, can be thought of as a crossing-symmetric  relativistic generalization of potential theory where Regge behavior holds in both s- and t-channels. (For a historical backdrop leading to the development of the Regge theory, see \cite{Chew,Gribov,DTFT}.)

Standard Regge analysis in the s-channel, where $s\rightarrow \infty$ with $t$ fixed, begins with the $t$-channel partial wave expansion
\be
A(s,t) =\sum_{J=0}^\infty (2J+1) a_J(t) P_J(z_t)
\ee
where $P_J(z_t)$ are the Legendre polynomials, with $z_t$ the cosine of the t-channel CM angle, $z_t=\cos \theta_t$. For identical scalars with mass $m$, $\cos \theta_t= 1+s/(t-4m^2)$, and  the amplitude,  as a function of $z_t$ with $t$ fixed, is even in $z_t$, i.e.,  the sum only involves even values of $J$. 
  One can re-write the partial wave expansion as a contour integral in the complex $J$-plane
\be
A(s,t) =- \int_{{\cal C}} \frac {dJ}{2 i} \frac {2J+1}{\sin \pi J} a(t,J) P_J(-z_t)
\ee
where the contour ${\cal C}$ encircles clock-wisely all non-negative integers. Here one has introduced an analytically continued partial-wave amplitude $a(t,J)$, which interpolates the physical partial wave amplitudes, $a_J(t)$, for integral $J$ values. This re-write, known as the Sommefeld-Watson transform, is simply a mathematical identity.

The key dynamical assumption involves the existence of a unique amplitude $a(t,J)$ which vanishes sufficiently rapidly for ${\rm Re} \; J\rightarrow +\infty$ so that one can open up the $J$-contour, with the contour at infinity dropped, leaving a contour running vertical along an imaginary axis~\footnote{This  interpolation with  physical partial-wave amplitudes should be done for even and odd $J$ separately, leading to even and odd ``signatured" amplitudes, $a^{\pm}(t,J)$.  For the present analysis, we only need to deal with the even-signatured amplitude, $a^{+}(t,J)$. }.
Using analyticity in $s$ and polynomial-boundedness, one can show such an amplitude always exists for ${\rm Re} J$ large and can be expressed  in a ``Froissart-Gribov" representation
\be
a(t,J)= \frac{2}{\pi} \int^\infty _{z_0}dz_t Q_J(z_t) {\rm Im} A(s,t)\; .
\ee
Here $Q_J(z_t)$ is the Legendre function of the second kind, $Q_J \sim a_0z_t^{-(J+1) }  + a_2 z_t^{-(J+3)} + \cdots $ for $z_t$ large,  $z_0=z_t(s_0,t)>1$, $s_0$ being the lowest s-channel threshold singularity, and $2i {\rm Im} A(s,t)$ is the discontinuity across this s-channel cut. Note that, for $s$ large, with $t$ fixed,
\be
 z_t \simeq s/(t-4m^2)\;.
\ee
Regge theory amounts to the assertion that the Froissart-Gribov formula interpolates all physical partial-wave amplitudes. As in potential theories, this analytically continued  amplitude $a(t,J)$ is expected to have $t$-dependent poles in $J$, i.e.,
\be
a(t,J)\sim  \frac{\beta(t)}{J-\alpha(t)}\; .
\ee
When $\alpha(t)$ reaches an integer $n$ at $t_n$, i.e., $\alpha(t_n)=n$, one can show, by reversing the Sommerfeld-Watson transform, a pole exists in $A(s,t)$ at $t=t_n$, corresponding to a spin-$n$ bound-state. One also notes that the existence of such a pole corresponds to the divergence of the Froissart-Gribov integral at large $z_t$, at fixed $t$.

One next pull this contour to the right, with the ${\rm Re}  J= -1/2$. In so doing, one picks up all Regge trajectories with ${\rm Re} \; \alpha(t)>-1/2$, leading to a Regge representation
\be
A(s,t) =- \int^{-1/2+i\infty}_{-1/2-i\infty} \frac {dJ}{2 i} \frac {1+e^{-i\pi J}}{2\sin \pi J} a(t,J) P_J(-z_t)   +\pi   \sum_k   \frac {1+e^{-i\pi \alpha_k}}{2\sin \pi \alpha_k} \beta_k(t)  P_{\alpha_k} (-z_t) \;.
\ee
Here we have absorbed the factor $2J+1$ and have also introduced a signature factor, $\frac {1+e^{-i\pi J}}{2}$, which projects out the sum over  even partial waves. Note that, for $s$ large, with $t$ fixed,   $\cos z_t \sim s$, 
thus leading to the Regge pole dominance by the leading Regge trajectory, e.g.,  (\ref{eq:4string2}) and (\ref{eq:Regge}).

\def\theequation{B.\arabic{equation}}
\setcounter{equation}{0}

\section{BCFT Shift,  Dressing Factors, and  Large $z$ Limit:}
We demonstrate explicitly how the dressing factors work by applying their $O(2,2)$ to the four-point Parke-Taylor amplitudes.  Consider first the standard BCFW shift for an adjacent  pair of vectors, $(1,2)$ 
\bea
|1\rangle&\rightarrow& |1'\rangle =  |1\rangle \nn
|2\rangle &\rightarrow & |2'\rangle =  |2 \rangle  - z|1\rangle  \; .
\eea
The shifted BCFW-amplitudes, $A_{(h_1',h_2',h_3,h_4)}(z;s_0,t)$,  obtained by the substitution of  $|1\rangle$ and $|2\rangle$ by $|1'\rangle$ and $|2'\rangle$ respectively  in the Parke-Taylor formula are
\bea
A_{(+',-',-,+)}(z;s_0,t)&=&\frac{\langle 2'3\rangle^4}{\langle 1'2'\rangle\langle 2'3\rangle\langle 34\rangle\langle 41'\rangle}=(1- z \langle 13\rangle/\langle 23\rangle)^3  A_{(+,-,-,+)}(s_0,t)   \rightarrow z^3\nn
A_{(-',+',-,+)}(z;s_0,t)&=&\frac{\langle 34\rangle^4}{\langle 1'2'\rangle\langle 2'3\rangle\langle 34\rangle\langle 41'\rangle}=\frac{1}{(1-z \langle 13\rangle/\langle 23\rangle)}A_{(-,+,-,+)}(s_0,t) \rightarrow z^{-1}\nn
A_{(-',-',+,+)}(z;s_0,t)&=&\frac{\langle 1'2'\rangle^4}{\langle 1'2'\rangle\langle 2'3\rangle\langle 34\rangle\langle 41'\rangle}=\frac{1}{(1-z \langle 13\rangle/\langle 23\rangle)}A_{(-,-,+,+)}(s_0,t) \rightarrow z^{-1}\nn
A_{(+',-',+,-)}(z;s_0,t)&=&\frac{\langle 2'4\rangle^4}{\langle 1'2'\rangle\langle 2'3\rangle\langle 34\rangle\langle 41'\rangle}=\frac{ (1- z \langle 14\rangle/\langle 24\rangle)^4}{(1-z \langle 13\rangle/\langle 23\rangle)    }   A_{(+,-,+,-)}(s_0,t)  \rightarrow z^3\nn
A_{(-',+',+,-)}(z;s_0,t)&=&\frac{\langle 1'4\rangle^4}{\langle 1'2'\rangle\langle 2'3\rangle\langle 34\rangle\langle 41'\rangle} = \frac{1}{(1-z \langle 13\rangle/\langle 23\rangle)}A_{(-,+,+,-)}(s_0,t) \rightarrow z^{-1}\nn
A_{(+',+',-,-)}(z;s_0,t)&=&\frac{\langle 34\rangle^4}{\langle 1'2'\rangle\langle 2'3\rangle\langle 34\rangle\langle 41'\rangle}=\frac{1}{(1-z \langle 13\rangle/\langle 23\rangle)}A_{(+,+,-,-)}(s_0,t) \rightarrow z^{-1}\; . \nn
\label{eq:BCFW-adj}
\eea

Let's now apply our modified shift,   where
\be
A_{physical}(\hat 1 \dots, \hat j\dots) = \alpha(z)^{-2h_1} \beta(z)^{2h_j}A_{BCFW}(\hat{1}\dots\hat{j}\dots)\; .
\ee
Since $\alpha(z)^2\sim \beta(z)^2 \rightarrow z$ for $z$ large, as shown in Table-\ref{tab:large-z}, large-z behavior of the modified amplitudes now agree with Regge expectation.
For example, one finds that, from Eq. (\ref{eq:BCFW-adj}),  
\bea
A^{physical}_{(+',-',-,+)}&=& (\alpha\beta)^{-2} A_{(+',-',-,+)}(z;s_0,t)= \frac{(1- z \langle 13\rangle/\langle 23\rangle)^3}{ \alpha(z)^2 \beta(z)^2}    A_{(+,-,-,+)}(s_0,t) \rightarrow z\nn
A^{physical}_{(-',+',-,+)}&=& (\alpha\beta)^{2} A_{(-',+',-,+)}(z;s_0,t)= \frac{\alpha(z)^2 \beta(z)^2 }{(1-z \langle 13\rangle/\langle 23\rangle)} A_{(-,+,-,+)}(s_0,t) \rightarrow z\nn
A^{physical}_{(-',-',+,+)}&=&A_{(-',-',+,+)}(z;s_0,t) =\frac{1}{(1-z \langle 13\rangle/\langle 23\rangle)}A_{(-,-,+,+)}(s_0,t) \rightarrow z^{-1}\; .
\label{eq:dressed-adj}
\eea
Similar expressions can easily be written down for $A_{(+',-',+,-)}(z;s_0,t)$, $A_{(-',+',-,+)}(z;s_0,t)$, and $A_{(-',-',+,+)}(z;s_0,t)$. These large-$z$ limits are again in agreement with that listed under column 2 in Table-\ref{tab:large-z}. 

Let us next turn to the evaluation of these amplitudes under  $O(2,2)$ signature where $\alpha(z)$ and $\beta(z)$ take on simpler form, given by (\ref{eq:alphabeta2}),
\be
\alpha(z)=\beta(z)=\sqrt{1-iz} = \sqrt{s(z)/s_0}\; .
\ee
One finds that, from Eq. (\ref{eq:BCFW-adj}),  
\bea
A^{physical}_{(+',-',-,+)}&=& (\alpha\beta)^{-2} A_{(+',-',-,+)}(z;s_0,t)= \frac{s(z)}{s_0}    A_{(+,-,-,+)}(s_0,t) = A_{(+,-,-,+)}(s(z),t)   \rightarrow z\nn
A^{physical}_{(-',+',-,+)}&=& (\alpha\beta)^{2} A_{(-',+',-,+)}(z;s_0,t)= \frac{s(z) }{s_0} A_{(-,+,-,+)}(s_0,t)=    A_{(-,+,-,+)}(s(z) ,t)  \rightarrow z\nn
A^{physical}_{(-',-',+,+)}&=&A_{(-',-',+,+)}(z;s_0,t) =\frac{s_0}{s(z)}A_{(-,-,+,+)}(s_0,t)=  A_{(-,-,+,+)}(s(z),t)   \rightarrow z^{-1}\; .\nn
\eea
That is, with the dressing factors, one not only reproduce the desired large-$z$ behavior, one also correctly restore the kinematic zeros in the finite $z$-plane.

This  can also be applied to all other helicity configurations, i.e., for ${(+',-',+,-)}$, ${(-',+',-,+)}$, and ${(-',-',+,+)})$, and also for non-adjacent shift.  For all cases,  the large-$z$ behavior for the resulting physical amplitudes is now in agreement with the Regge expectation.  However, we note that, for $A^{physical}_{(+',-',+,-)}$ and $A^{physical}_{(-',+',-,+)}$, they agree with the exact Parke-Taylor amplitudes except for a   factor $(s(z)/u(z))^2$ and $(u(z)/s(z))^2$ respectively, which both approach 1 for $z$-large with $t$ fixed. It is unclear to us what is the cause of this discrepancy, which could be due to the inadequacy of directly applying the Parke-Taylor formula under the $O(2,2)$ signature.  The source of this additional dressing factor is being studied.

\def\theequation{C.\arabic{equation}}
\setcounter{equation}{0}

\section{Generic helicity configurations and large-$z$ behavior prescribed by CSW rules}

In this appendix we show that for generic helicity configuration the
asymptotic behavior under momentum shifting does agree with Regge
expectation demonstrated in Table 1. For the purpose of discussion
let us consider Cachazo-Svrcek-Witten (CSW) construction\cite{Cachazo:2004kj}
of Yang-Mills amplitudes. In particular we note that the CSW rules
can be made manifest through canonical transformation of field variables
in light-cone gauge action\cite{Mansfield:2005yd}, see also \cite{Ettle:2006bw,Fu:2009cg}.
The transformed action prescribes scalar propagator $1/p^{2}$ and
MHV vertices explicitly in the form of Parke-Taylor formula, while
off-shell continuation is taken so that all spinors in the formula
are defined through light-cone {\it coordinate components} described 
exactly as those given in  Eq. (\ref{eq:spinors}).
\begin{figure}[h!]
\centering
\subfigure[]{
\includegraphics[width=2.5cm]{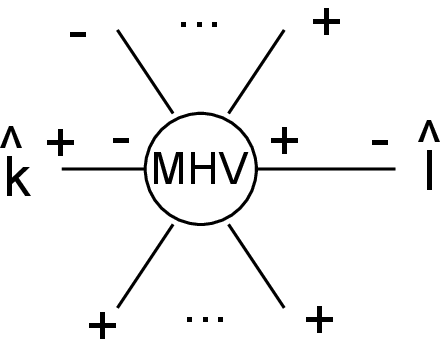}
\label{leading}
}
\hspace{0.5cm}
\subfigure[]{
\includegraphics[width=4.5cm]{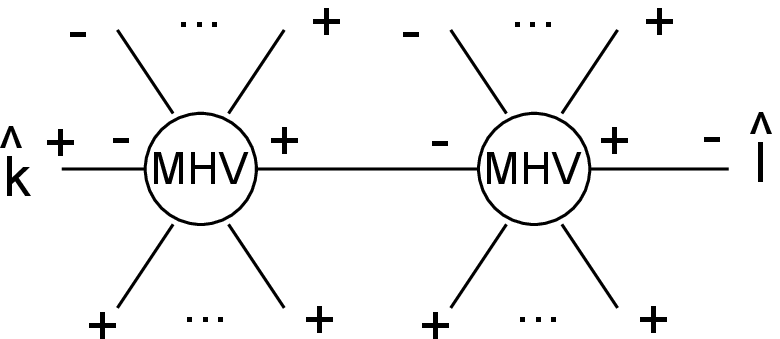}
\label{subleading}
}
\caption{Leading  and subleading  MHV graphs as $z \rightarrow \infty$}
\end{figure}
Generically in a shifted amplitude the shifted pair $(i^{\pm},j^{\pm})$
is connected by a string of propagators with MHV vertices inserted
between. See Fig. \ref{leading} and \ref{subleading} as illustrations, 
whereas subtrees can
be attached to unshifted legs. It is straightforward to see
that graphs that have more than one vertex connecting $(i^{\pm},j^{\pm})$
are subleading, so the calculation of large-$z$ asymptotic for generic
helicity configurations boils down to calculation of large-$z$ asymptotic
of a single shifted Parke-Taylor formula (multiplied by unshifted
subtrees which contain the momentum dependence of rest of the legs)
and therefore agrees with the MHV analysis discuss previously.

\bibliographystyle{utphys}
\bibliography{ReggeBCFW}


\end{document}